\documentclass[12pt]{iopart}
\usepackage[english]{babel}
\usepackage{graphicx}

\begin{document}
\title[Random sequential adsorption of tetramers]{Random sequential adsorption of tetramers}
\author{Micha\l{} Cie\'sla}
\address{M. Smoluchowski Institute of Physics, Jagiellonian University, 30-059 Kraków, Reymonta 4, Poland}
\ead{michal.ciesla@uj.edu.pl}
\begin{abstract}
Adsorption of tetramer built of four identical spheres was studied numerically using the Random Sequential Adsorption (RSA) algorithm. Tetramers were adsorbed on a two dimensional, flat and homogeneous surface. Two different models of the adsorbate were investigated: a rhomboid and a square one; monomer centres were put on vertices of rhomboids and squares, respectively. Numerical simulations allow to establish the maximal random coverage ratio as well as the Available Surface Function (ASF), which is crucial for determining kinetics of the adsorption process. These results were compared with data obtained experimentally for KfrA plasmid adsorption. Additionally, the density autocorrelation function was measured.
\end{abstract}
\pacs{68.43.Fg 05.45.Df}
\maketitle
\section{Introduction}
Irreversible adsorption is of major significance for many fields such as medicine and material sciences as well as pharmaceutical and cosmetic industries. For example, adsorption of some proteins is crucial for blood coagulation, inflammatory response, fouling of contact lenses, plaque formation, ultrafiltration and the operation of membrane filtration units. Moreover, controlled adsorption is fundamental for efficient chromatographic separation and purification, gel electrophoresis, filtration, as well as the performance of bioreactors, biosensing and immunological assays.
\par
Random Sequential Adsorption (RSA),  since its introduction by Feder \cite{bib:Feder1980}, has became a well established method used in numerical modelling of irreversible adsorption of, at first spherical molecules, and then of complex ones such us polymers or proteins, e.g. \cite{bib:Talbot1989, bib:Vigil1989, bib:Tarjus1991, bib:Viot1992, bib:Ricci1992}. Recent studies show that, for the purposes of adsorption modelling, complex molecules can be successfully approximated using coarse-grain models \cite{bib:Rabe2011, bib:Finch2012, bib:Katira2012, bib:Adamczyk2012rev}. For example, a coarse-grain model of fibrinogen can successfully explain the density of adsorbed monolayer for a wide range of experimental conditions \cite{bib:Adamczyk2010, bib:Adamczyk2011, bib:Ciesla2013-fib}. 
\par
This study focuses on the RSA of tetramers on a flat and homogeneous two dimensional collector surface. The interest in this topic was aroused by the work of Adamczyk et al. \cite{bib:Adamczyk2012}, who had experimentally measured adsorption of a KfrA plasmid and shown that it aggregates to a tetramer during adsorption. However, the tetramer structure is common for a number of substrates. Some of the latest theoretical studies in this field investigate Ag tetramer \cite{bib:Mazheika2011} and melanin \cite{bib:Orivea2012}; however, by using the density-functional approach. On the other hand, in similar works involving RSA modelling, dimer \cite{bib:Ciesla2012a, bib:Ciesla2013a} and polymer adsorption \cite{bib:Ciesla2013c} has been studied. The primary aim of this paper is to find the maximal random coverage ratio of monolayers built as a result of the irreversible tetramer adsorption. Additionally, I want to calculate parameters needed to estimate the kinetics of the adsorption process.
\section{Model}
\label{sec:Model}
\begin{figure}[htb]
\centerline{%
\includegraphics[width=6cm]{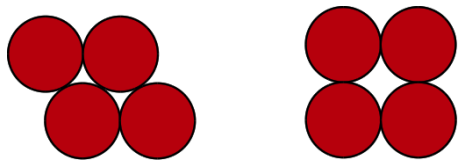}}
\caption{Two approximations of a tetramer used in simulations. All the monomers have radius $r_0$. Centres of monomers form a rhomboid (left) or a square (right).}
\label{fig:model}
\end{figure}

In this study two independent models of a tetramer are considered. In the first one, spheres of radius $r_0$, which represent monomers, form a rhomboid (rhomboid model) whereas in the second one, these spheres are placed at vertices of a square (square model); see Fig.\ref{fig:model}.
\par 
Modelled molecules are placed on a flat homogeneous collector surface according to the Random Sequential Adsorption (RSA) algorithm \cite{bib:Feder1980}, which iteratively repeats the following steps:
\begin{description}
\item[1.] a virtual tetramer is created. Its position and orientation on a collector is chosen randomly according to the uniform probability distribution; however, centres of all the components (see Fig.\ref{fig:model}) are required to be on a collector; 
\item[2.] an overlapping test is performed for previously adsorbed nearest neighbours of the virtual molecule. The test checks if surface-to-surface distance between each of the spheres is greater than zero;
\item[3.] if there is no overlap the virtual molecule is irreversibly adsorbed and added to an existing covering layer. Its position does not change during further calculations;
\item[4.] if there is an overlap, the virtual tetramer is removed and abandoned.
\end{description}
The number of RSA iterations $N$ is typically expressed in dimensionless time units:
\begin{equation}
\label{eq:dimlesstime}
t = N\frac{S_{\rm M}}{S_{\rm C}},
\end{equation}
where $S_{\rm M} = 4\pi r_0^2$ is an area covered by a single tetramer and $S_{\rm C}$ is a collector size. In case of these simulations, square collectors were used with a side size of $1000r_0$, so $S_{\rm C} = 10^6 r_0^2$, and algorithm was stopped after $t = 10^5$.
\par
\begin{figure}[htb]
\vspace{1cm}
\centerline{%
\includegraphics[width=6cm]{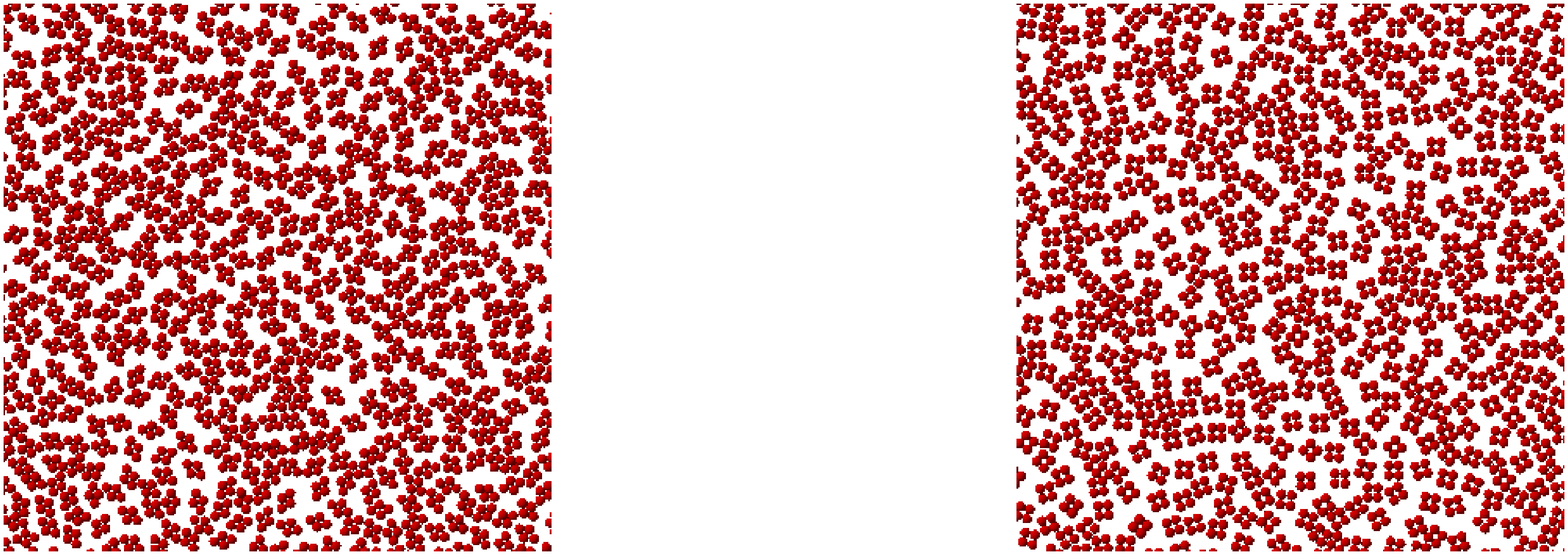}}
\caption{Example coverages for two different tetramer models: a rhomboid model (left) and a square model (right).}
\label{fig:examples}
\end{figure}

Obtained example coverages are presented in Fig.\ref{fig:examples}.
\par
The main parameter observed during simulation was a coverage ratio $\theta$:
\begin{equation}
\theta(t) = n(t) \frac{S_{\rm M}}{S_{\rm C}},
\end{equation}
where $n(t)$ is a number of adsorbed molecules after time $t$. To decrease statistical error, $100$ independent RSA simulations were performed for a single model.
\section{Results and discussion}
The main property of an adsorption layer is its maximal coverage ratio - the area covered by  particles adsorbed after a long enough period of time. Moreover, the adsorption kinetics can also be measured; however, it depends not only on the properties of collector and adsorbate particles, but also on the transport process, which brings those particles to surface proximity. On the other hand, due to a finite time of the RSA simulation, kinetics of the process has to be known to get appropriate values of $\theta_{\rm max} \equiv \theta(t\to\infty)$. Therefore, both the maximal coverage ratio and adsorption kinetics should be analysed together.
\subsection{Maximal random coverage ratio}
The kinetics of the RSA of spheres obeys the Feder law \cite{bib:Swendsen1981, bib:Privman1991}:
\begin{equation}
\theta_{\rm max} - \theta(t) \sim t^{-1/d},
\label{fl}
\end{equation}
where $d$ is a collector dimension and $t$ is a dimensionless time (\ref{eq:dimlesstime}). The relation (\ref{fl}) has been proved numerically for a one to six dimensional space~\cite{bib:Torquato2006} and also for fractal collectors, having $d<2$~\cite{bib:Ciesla2012b}, as well as for $2<d<3$~\cite{bib:Ciesla2013b}. Feder's law appeared to be valid also for different adsorbates like dimers~\cite{bib:Ciesla2012a} and polymers \cite{bib:Ciesla2013c}, however, for highly anisotropic molecules, parameter $d$ grows with a number of degrees of freedom of adsorbate particle \cite{bib:Ciesla2013c, bib:Hinrichsen1986}.
\par
\begin{figure}[htb]
\vspace{1cm}
\centerline{%
\includegraphics[width=6cm]{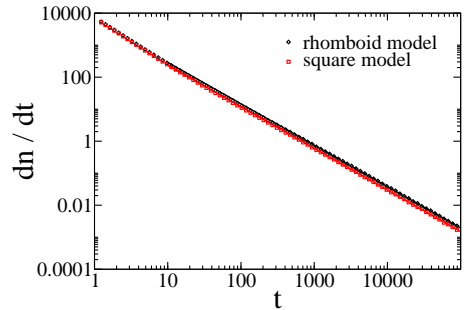}}
\caption{The dependence of the mean number of adsorbed particles on the dimensionless time. Diamonds and squares are simulation data for rhomboid and square model, respectively, whereas solid lines correspond to power fits obtained for $t>1000$: $dn / dt = 5714.1  t^{-1.294}$ for rhomboid model  and  $dn / dt = 4693.9  t^{-1.299}$ for square model. Corresponding values of exponent $d$ in (\ref{fl}) are $d=3.40$ and $d=3.34$ for rhomboid and square model, respectively.}
\label{fig:dnt}
\end{figure}

For large enough time $t$, the exponent can be measured directly from $d\theta / dt$ dependence on $t$ using the least squares approximation method (see Fig.\ref{fig:dnt}). 
It is interesting, that the obtained values of parameter $d$ for both the models are significantly higher than $2$, which is expected value for spherical particles. It suggests that orientational degree of freedom of the model of plasmid aggregate cannot be neglected, even though the shape anisotropy in this case is quite small.
\par
Having determined the exponent $d$ let $y=t^{-1/d}$. Then Eq.(\ref{fl}) follows to $\theta(y) = \theta_{\rm max} - A y$, where $A$ is a constant coefficient. Approximation of this linear relation for $y=0$ reveals the maximal random coverage $\theta_{\rm max}=0.5214$ and $\theta_{\rm max}=0.4910$ for the rhomboid and square model, respectively. The relative error of both values is approx. $0.5\%$. The difference is significant but relatively small and therefore it could not be measured experimentally, as the typical error of experimental methods is at the level of $5\%$. In both the cases, the maximal random coverage ratio is smaller than $\theta_{\rm max} \approx 0.54$ obtained for spheres \cite{bib:Torquato2006}, dimers \cite{bib:Ciesla2012a} or very short polymers \cite{bib:Ciesla2013c}.
\par
Adsorption of tetramers has been measured experimentally by Adamczyk et al.\cite{bib:Adamczyk2012}; they studied adsorption of KfrA plasmid on latex particles. KfrA plasmid is a spherical $39.22$ kDa particle and with a diameter of $4.5$ nm. However, the study has shown that KfrA aggregates. The size of the aggregate indicates that it contains four plasmids. AFM observation suggest that the aggregate can be described by the rhomboid model. Obtained values of maximal coverage ratio corresponds to surface density of $2.0-2.1$ mg/m$^2$ of KfrA, and is almost twice higher than measured experimentally. However, KfrA plasmids have uncompensated charge of 12 $e$ \cite{bib:Adamczyk2012}, which significantly lowers the coverage density due to electrostatic repulsion \cite{bib:Ciesla2013a}. Using effective equation \cite{bib:AdamczykBook}:
\begin{equation}
\theta_{\rm eff} = \frac{\theta_{\rm max}}{(1+H)^2},
\end{equation}
where $\theta_{\rm eff}$ is measured coverage and $H$ is the effective interaction range characterising the repulsive double-layer interaction and depends on electrostatic properties of a particles as well as dielectric properties of the solution, one can find the value of $H$. In this case, it is $0.39$ for the rhomboid model and $0.35$ for the square one.
\subsection{Adsorption kinetics}
\begin{figure}[htb]
\vspace{1cm}
\centerline{%
\includegraphics[width=6cm]{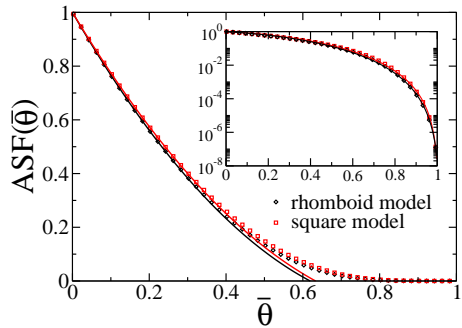}}
\caption{The dependence of Available Surface Function on a normalised coverage ratio. Diamonds and squares are simulation data for the rhomboid and square model, respectively. Solid lines correspond to the fits (\ref{eq:asffit}) obtained for $\bar{\theta}<0.2$: $ASF(\theta) = 1 - 4.741 \theta + 5.09 \theta^2$ for the rhomboid model  and  $ASF(\theta) = 1 - 4.839\theta + 5.218\theta^2$ for the square one.}
\label{fig:asf}
\end{figure}

Adsorption kinetics is governed by two factors. The first one is transport process shifting molecules to surface proximity. As it depends on a given experiment's conditions, it hardly enters the general theoretical analysis. The second factor is probability of adsorption; it decreases in time due to diminishing area of uncovered collector surface. The dependence between adsorption probability and temporary coverage ratio is known as Available Surface Function (ASF) and it can be easily determined from the RSA simulation. Its dependence on normalised coverage $\bar{\theta} = \theta / \theta_{\rm max}$ is shown in Fig.\ref{fig:asf}.
At the limit of low coverage, the $ASF(\theta)$ is commonly approximated by a quadratic fit \cite{bib:Ricci1992, bib:Tarjus1991, bib:AdamczykBook}:
\begin{equation}
ASF(\theta) = 1 - C_1 \theta + C_2 \theta^2 + O(\theta^3).
\label{eq:asffit}
\end{equation}
The expansion coefficients $C_1$ corresponds to the surface area blocked by a single tetramer, whereas $C_2$ denotes a cross-section of the area blocked by two independent molecules. Both of them have major significance because they are directly related to the second $B_2=1/2C_1$ and third $B_3=1/3C_1^2-2/3C_2$ viral coefficients of the equilibrium tetramer monolayer \cite{bib:Tarjus1991,  bib:AdamczykBook}. For example, the 2D pressure $P$ and the chemical potential of tetramer $\mu$ can be expressed via the series expansion at a low coverage limit \cite{bib:AdamczykBook}
\begin{equation}
\begin{array}{c}
P = \frac{k_{\rm B} T}{S_{\rm F}} \left( \theta + B_2\theta^2 + B_3\theta^3 + o(\theta^3) \right), \\
\mu = \mu_0 + k_{\rm B} T \left( \ln \theta  + 2B_2\theta + \frac{3}{2}B_3 \theta^3 + o(\theta^3)   \right), 
\end{array}
\end{equation}
where $k_B$ is the Boltzmann constant, $T$ is the absolute temperature,
and $\mu_0$ is the reference potential.
\par
Results presented in Fig.\ref{fig:asf} show that $C_1$ is slightly bigger for the square model, which is expected as the overall size of the particle in this model is slightly bigger than in the rhomboid one. It is also worth to notice that the mean surface blocked by a tetramer is approximately $20\%$ larger than for a spherical particle, for which $C_1=4$. On the other hand, in case of tetramers $C_2>5$, whereas for spheres $C_2 \approx 3.308$. Therefore, due to opposite signs at $C_1$ and $C_2$ in Eq.\ref{eq:asffit}, the ASF difference between spheres and tetramers is getting smaller for slightly larger coverages, when the parameter $C_2$ becomes more important.
\par
\begin{figure}[htb]
\vspace{1cm}
\centerline{%
\includegraphics[width=6cm]{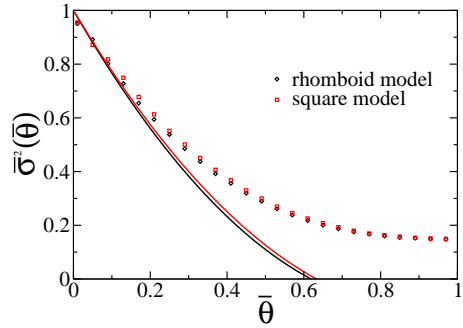}}
\caption{The ASF approximation by the normalised density variance $\bar{\sigma}^2$. Diamonds and squares are measured values  for rhomboid and square model, respectively; whereas solid lines correspond to the ASF fit in a low coverage limit (\ref{eq:asffit}).}
\label{fig:varasf}
\end{figure}

Note that in a limit of small coverages ${\rm ASF}(\theta)$ can also be estimated experimentally. It has been shown that in these conditions ${\rm ASF}(\theta) = \bar{\sigma}^2(\theta)$, where $\bar{\sigma}^2(\theta) = \sigma^2(\theta) / \langle n(\theta) \rangle$ is a normalised variance of the number of adsorbed particles \cite{bib:Schaaf1995}. The typical experimental procedure used for estimation of $\bar{\sigma}^2(\theta)$, described in \cite{bib:Adamczyk1996}, can be used also for monolayers generated by the RSA. Fig.\ref{fig:varasf} shows results of such calculations for tetramer layers in comparison with theoretical fit (\ref{eq:asffit}) obtained above.
As expected, agreement is good only for $\bar{\theta}<0.2$.
\par
At the jamming limit, the ASF for anisotropic molecules is typically approximated by \cite{bib:Ricci1992}:
\begin{equation}
\label{asfjfit}
{\rm ASF}(\bar{\theta}) = (1 + a_1 \bar{\theta} + a_2 \bar{\theta}^2 + a_3 \bar{\theta}^3)(1-\bar{\theta})^4
\end{equation} 
As shown in Fig.\ref{fig:asf} inset, the above relation is also valid for tetramer adsorption. The fit can be directly used for finding adsorption kinetics when transport is provided by diffusion or convection. Details of this procedure have been described elsewhere \cite{bib:Adamczyk2010, bib:Ciesla2013a}.
\subsection{Density autocorrelation}
\begin{figure}[htb]
\vspace{1cm}
\centerline{%
\includegraphics[width=6cm]{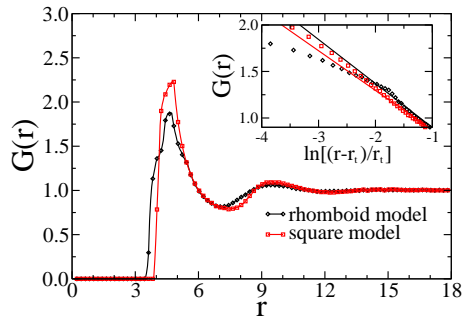}}
\caption{Density autocorrelation function G(r) for rhomboid and square models. Inset shows a logarithmic singularity at $r \to r_t+$. Parameter $r_t=4.68$ and $t_t=4.78$ for the rhomboid and square model, respectively.}
\label{fig:cor}
\end{figure}

The Density autocorrelation function gives additional insight into coverages structure. It is defined as:
\begin{equation}
G(r) = \frac{P(r)}{2\pi r \rho},
\end{equation}
where $P(r)dr$ is a probability of finding two tetramers in a distance between $r$ and $r+dr$. Here, the distance $r$ is measured between the geometric centres of tetramers.  Parameter $\rho$ is the mean density of particles inside a covering layer. Such a normalisation leads to $G(r\to \infty) = 1$. In the case of spherical particles, $G(r)$ has a logarithmic singularity in the touching limit \cite{bib:Swendsen1981} and superexponential decay at large distances \cite{bib:Bonnier1994}. The density autocorrelation function is shown in Fig.\ref{fig:cor}.
It can be noticed that the autocorrelation density for the rhomboid model is slightly shifted left compared to the one for the square model. This is due to slightly denser packing of monomers in the rhomboid model and, hence, a somewhat smaller distance possible between the particles. The first maximum is smaller because rhomboid shape anisotropy is bigger than the square one, which results in broader dispersion of possible distances. Due to anisotropy of shapes in both models used, there is no singularity at small $r$. However, the $G(r)$ slope past the first maximum resembles logarithmic singularity, especially for the square model, which is shown in the Fig.\ref{fig:cor} inset. For large $r$, autocorrelations approaches their limit value very fast, which is similar to the case of spheres maximal random coverages.
\section{Summary}
The maximal random coverage ratio of a tetramer monolayer is $\theta_{\rm max}=0.5214$ and $\theta_{\rm max}=0.4910$ for the rhomboid and square model, respectively. In both cases, the ratio is slightly smaller than for spheres. On the other hand, in a limit of low coverage density, the surface is filled approximately $20\%$ faster by tetramers than by equally sized spheres. At jamming limit, RSA of tetramers shows behaviour typical to anisotropic molecules, which is rather unexpected considering small shape anisotropy of both models used. Properties of the density autocorrelation function are similar to those of the case of spheres.
\section*{Acknowledgement}
This work was supported by grant MNiSW/N N204 439040.

\end{document}